\documentclass[twocolumn,prb,reprint,superscriptaddress,floatfix,amsmath,amssymb,aps]{revtex4-2}
\pdfoutput=1

\usepackage[utf8]{inputenc}
\usepackage[caption=false,subrefformat=parens,labelformat=parens]{subfig}
\usepackage{graphicx}
\usepackage{float}
\usepackage{amsmath}
\usepackage{dcolumn}
\usepackage{bm}
\usepackage{xcolor}
\usepackage{xfrac}
\usepackage[colorlinks=true, allcolors=blue]{hyperref}
\usepackage{chemformula}
\usepackage{siunitx}

\begin{document}
\title{Pseudogap behavior in charge density wave kagome material ScV$_6$Sn$_6$ revealed by magnetotransport measurements}

\author{Jonathan M. DeStefano}
\affiliation{Department of Physics, University of Washington, Seattle, WA 98112, USA}
\author{Elliott Rosenberg}
\affiliation{Department of Physics, University of Washington, Seattle, WA 98112, USA}
\author{Olivia Peek}
\affiliation{Department of Physics, University of Washington, Seattle, WA 98112, USA}
\author{Yongbin Lee}
\affiliation{Ames Laboratory, U.S. Department of Energy, Ames, Iowa 50011, USA}
\author{Zhaoyu Liu}
\affiliation{Department of Physics, University of Washington, Seattle, WA 98112, USA}
\author{Qianni Jiang}
\affiliation{Department of Physics, University of Washington, Seattle, WA 98112, USA}
\author{Liqin Ke}
\affiliation{Ames Laboratory, U.S. Department of Energy, Ames, Iowa 50011, USA}
\author{Jiun-Haw Chu}
\affiliation{Department of Physics, University of Washington, Seattle, WA 98112, USA}

\date{\today}

\begin{abstract}
Over the last few years, significant attention has been devoted to studying the kagome materials \textit{A}\ch{V3Sb5} (\textit{A} = K, Rb, Cs) due to their unconventional superconductivity and charge density wave (CDW) ordering. Recently \ch{ScV6Sn6} was found to host a CDW below $\approx$\SI{90}{K}, and, like \textit{A}\ch{V3Sb5}, it contains a kagome lattice comprised only of V ions. Here we present a comprehensive magnetotransport study on \ch{ScV6Sn6}. We discovered several anomalous transport phenomena above the CDW ordering temperature, including insulating behavior in interlayer resistivity, a strongly temperature-dependent Hall coefficient, and violation of Kohler's rule. All these anomalies can be consistently explained by a progressive decrease in carrier densities with decreasing temperature, suggesting the formation of a pseudogap. Our findings suggest that high-temperature CDW fluctuations play a significant role in determining the normal state electronic properties of \ch{ScV6Sn6}.
\end{abstract}

\maketitle

\section{Introduction}


Materials containing kagome lattices have emerged as a promising platform for studying the interplay of electronic correlations and topology~\cite{Kuroda2017,Ye2018,kang2020,TMS_topologicalmagnetism}. Among these, kagome metals hosting charge density waves have gained significant attention due to their novel symmetry breaking phases and rich phase diagrams~\cite{FeGe_CDW, RVS_props, CVS_correlated}. The \textit{A}\ch{V3Sb5} (\textit{A} = K, Rb, Cs) family hosts a charge density wave (CDW) with $T_{CDW}$ at $\approx$\SI{80}{K}, which potentially breaks time reversal symmetry~\cite{TRSbreaking_kagome, lackofTRSbreaking_kagome} and rotational symmetry~\cite{CVS_twofold, Xu2022, Li2022}, leading to speculation about an orbital current loop state~\cite{fengChiralFluxPhase2021, Denner2021, Lin2021} and electronic nematicity~\cite{Nie2022}. At lower temperatures ($<$\SI{3}{K}), a superconducting state coexists and competes with the CDW~\cite{KVS_SC, RVS_props, CVS_correlated}, displaying signatures of a pair density wave~\cite{chenRotonPairDensity2021}. Many of these phenomena resemble the characteristics of other strongly correlated systems, such as high temperature superconductors, where the extended fluctuation regime gives rise to intertwined orders and complex phase diagrams~\cite{Intertwined}.

\ch{ScV6Sn6} is the latest addition to the set of kagome metals exhibiting CDWs, with a CDW transition temperature near \SI{90}{K}~\cite{SVSdiscovery}. Since this compound contains kagome layers comprised solely of V ions, it is natural to compare it to the \textit{A}\ch{V3Sb5} family. However, early studies have found several distinct differences between \ch{ScV6Sn6} and these compounds. In \ch{ScV6Sn6} the CDW is associated with a $\sqrt{3}$ x $\sqrt{3}$ in-plane ordering~\cite{SVSdiscovery}, which is different from the 2 x 2 ordering in \textit{A}\ch{V3Sb5} where the wave vectors nest the van Hove singularities of the kagome-derived energy bands. In \ch{ScV6Sn6} the lattice distortion associated with the CDW is mostly along the c-axis~\cite{SVSdiscovery} whereas the distortion in \textit{A}\ch{V3Sb5} is mostly in the ab-plane~\cite{AVS_224}. Unlike the \textit{A}\ch{V3Sb5} family, no superconductivity has been found in \ch{ScV6Sn6} down to \SI{40}{mK} even under high pressures~\cite{SVSpressure}. Nevertheless, similar to the \textit{A}\ch{V3Sb5} family, signatures of time reversal symmetry breaking have been suggested by muon spin relaxation rate measurements and an anomalous Hall effect~\cite{hiddenmagnetism_SVS, Mandrus_SVShall, SVS_AHE}.  Recent measurements, including scanning tunneling microscopy, angle-resolved photoemission spectroscopy, and Raman spectroscopy suggest the CDW is primarily structurally driven~\cite{SVS_STM,tuniz2023dynamics,SVS_raman}, indicating a minor role of the electronic degrees of freedom in the CDW formation. However, despite the first-order nature of the CDW transition, recent studies have revealed short-range CDW fluctuations persisting well above $T_{CDW}$ in \ch{ScV6Sn6}~\cite{competingCDW_SVS}. Hence, it is crucial to examine whether these fluctuations impact the electronic properties, as observed in other strongly correlated systems. 

In this paper, we present evidence of a pseudogap above the CDW transition in \ch{ScV6Sn6}. Pseudogap formation was first observed in the cuprate superconducting family, and it refers to the suppression of the density of states which was revealed by various spectroscopy measurements and anomalous transport behavior~\cite{Timusk_1999}. Our conclusion of pseudogap formation in \ch{ScV6Sn6} is established from a comprehensive magnetotransport study, including measurements of the interlayer resistivity, magnetoresistance, and the Hall effect, all consistent with an abnormal decrease of carrier density with decreasing temperature above the CDW transition. In addition, we found several striking similarities to the proposed pseudogap phase in the Fe-based superconductors, in which strong spin density wave fluctuations persist well above the transition temperature. Our results suggest that there is an extended fluctuation regime in \ch{ScV6Sn6} which strongly influences the electronic transport properties above the transition temperature.

\section{Results}

\subsection{Zero-field Resistivity}

\begin{figure}
    \centering
    \includegraphics[width=0.5\textwidth]{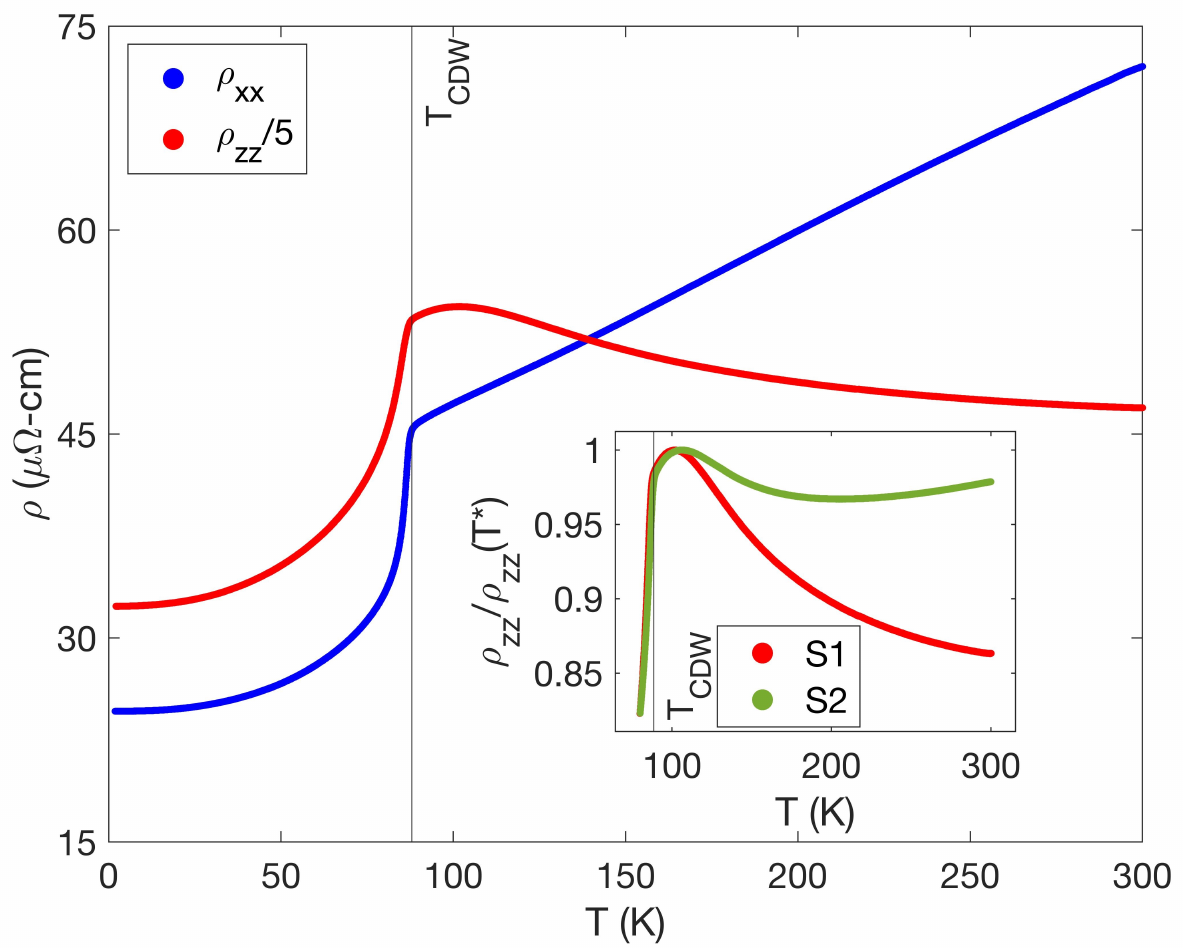}
    \caption{$\rho_{xx}$ and $\rho_{zz}$ (divided by 5) of \ch{ScV6Sn6} as a function of temperature. $T_{CDW}$ is marked with a vertical line. Inset: $\rho_{zz}$ of two different samples each normalized to be 1 at $T^*$ as discussed in the main text. $T_{CDW}$ is marked in the same way as the main figure.}
    \label{fig:RvT}
\end{figure}

Fig.~\ref{fig:RvT} presents in-plane resistivity, $\rho_{xx}$, and interlayer resistivity, $\rho_{zz}$ (divided by 5), of a typical \ch{ScV6Sn6} sample as a function of temperature while cooling. $\rho_{xx}$ is consistent with previous reports~\cite{SVSdiscovery, SVSpressure}, with residual resistivity ratios of samples ranging from 3-10. Drops in resistivity are present in both curves near \SI{90}{K}, indicating the charge density wave transition ($T_{CDW}$)~\cite{SVSdiscovery}. It should be noted that this is a first order transition, but only the cooling curve is shown due to the small temperature hysteresis of this transition ($\approx$1-\SI{2}{K}). As previously noted~\cite{SVSpressure}, charge density wave phase transitions in layered materials typically appear as hump-like increases in resistivity as a function of temperature, as parts of the Fermi surface get gapped out by the transition.

The drop of resistivity at $T_{CDW}$ is reminiscent of the spin density wave transition in the parent compounds of the iron pnictide superconductors, such as \ch{BaFe2As2}, which is also characterized by a similar feature~\cite{BaCo122_caxis}. This unusual behavior in \ch{BaFe2As2} is understood as a more rapid decrease of scattering rates relative to the decrease of carrier density below the spin density wave transition. Like observed in \ch{BaFe2As2}, optical measurements of \ch{ScV6Sn6}~\cite{SVSoptics} have also revealed a similar decrease in both the carrier density and the scattering rate below $T_{CDW}$, which could explain the increase in conductivity. The enhanced electron scattering above $T_{CDW}$ can be explained by the competing CDW fluctuations above $T_{CDW}$~\cite{competingCDW_SVS}.


The interlayer resistivity measurements reveal a moderate resistivity anisotropy in \ch{ScV6Sn6} with $\rho_{zz}$ roughly 5 times larger than $\rho_{xx}$ at \SI{2}{K}. This is considerably smaller than in \ch{CsV3Sb5} where $\rho_{zz}$ is $\approx$20 times larger than $\rho_{xx}$ at low temperatures~\cite{CVS_twofold}, implying \ch{ScV6Sn6} is more 3-dimensional than \ch{CsV3Sb5}. Nevertheless, unlike \ch{CsV3Sb5}, the temperature dependence of $\rho_{zz}$ is dramatically different from $\rho_{xx}$, showing a broad maximum roughly \SI{15}{K} above $T_{CDW}$ at $T^*$. As shown in the inset of Fig.~\ref{fig:RvT}, for temperatures above $T^*$ multiple \ch{ScV6Sn6} samples display insulating behavior. The difference in high temperature resistivity between samples may be explained by contamination from lower resistivity in-plane components.

The insulating interlayer resistivity and metallic in-plane resistivity have been observed in highly anisotropic layered materials such as \ch{Sr2RuO4}, in which the much weaker interlayer tunneling results in incoherent c-axis transport~\cite{Sr2RuO4_incoherenttransport}. However, such a phenomenon is usually seen in materials with resistivity anisotropy $\rho_{zz}/\rho_{xx} \gg 10$, which is not applicable to \ch{ScV6Sn6} where $\rho_{zz}/\rho_{xx} \approx 5$. Interestingly, a similar insulating temperature dependence with a broad maximum in $\rho_{zz}$ above a density wave transition has also been observed in \ch{BaFe2As2}~\cite{BaCo122_caxis, BFAP_caxis, BaFe2As2_bunchofdopings_caxis, BKFA_caxis, BFRA_caxis}, which also has a moderate resistivity anisotropy ($\rho_{zz}/\rho_{xx} \approx 7$). The insulating $\rho_{zz}$ in \ch{BaFe2As2} was interpreted as signature of a pseudogap, resulting from the spin density wave fluctuations partially gapping the section of the Fermi surface where the Fermi velocity has a large z-component. We propose that a similar mechanism could be responsible for the insulating $\rho_{zz}$ in \ch{ScV6Sn6}, which is further supported by the Hall effect and magnetoresistance analysis presented in the following sections.



\subsection{Hall Effect}
\label{subsection:Hall}

\begin{figure*}
    \centering
    \includegraphics[width=0.9\textwidth]{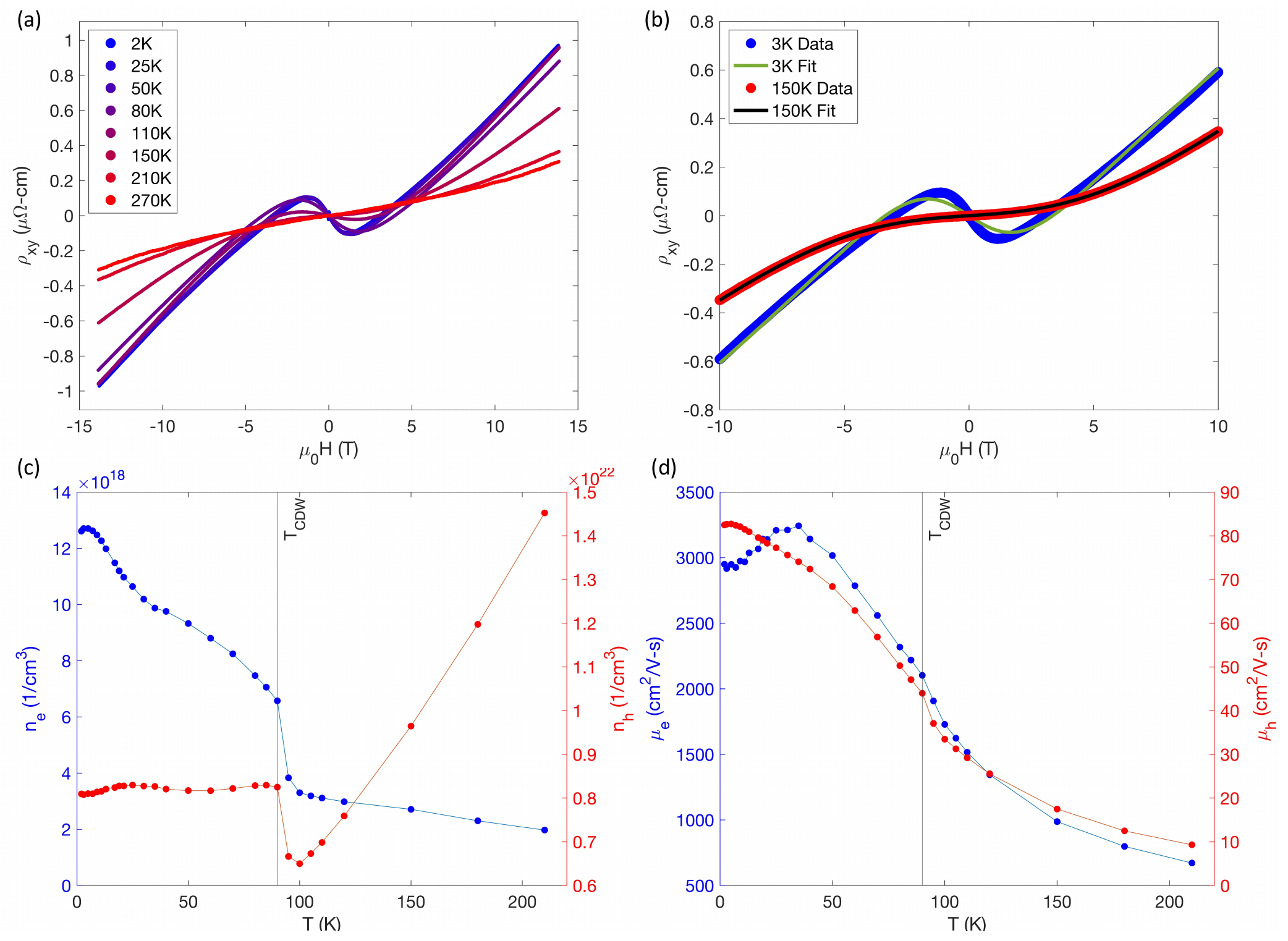}
    \caption{(a) $\rho_{xy}$ as a function of magnetic field at various temperatures. (b) $\rho_{xy}$ as a function of magnetic field at \SI{2}{K} and \SI{150}{K} with fits to the two-band model. (c) $n_e$ and $n_h$ as a function of temperature extracted from fits to the two-band model. (d) $\mu_e$ and $\mu_h$ as a function of temperature extracted from fits to the two-band model.}
    \label{fig:hall}
\end{figure*}

Fig.~\ref{fig:hall}(a) presents $\rho_{xy}$ as a function of magnetic field at a variety of temperatures. Across the entire measured temperature range $\rho_{xy}$ is non-linear, but while it evolves smoothly as a function of temperature above $T_{CDW}$, it stays relatively unchanged below $T_{CDW}$. There are two possible sources for the non-linearity in $\rho_{xy}$: the multi-band effect and the anomalous Hall effect. We first present an analysis of fitting of $\rho_{xy}$ using a two-band model, which reveals a strong temperature dependence of carrier density above the CDW transition. We will also argue that this conclusion can be made from the high field Hall coefficient of $\rho_{xy}$ even without any two-band fitting.    

In order to analyze the two-band Hall effect, the standard two-band model is used to simultaneously fit $\rho_{xx}(\mu_0H)$ and $\rho_{xy}(\mu_0H)$ at each temperature~\cite{CrO2_2bandhall}. The carrier densities and mobilities were determined by a non-linear least squares minimization of the error $\rho_{xy}-\rho^{fit}_{xy} +C(\rho_{xx}-\rho^{fit}_{xx})$ where C provided a weighting such that $\rho_{xy}$ was prioritized (as $\rho_{xx}$ has potentially more scattering contributions than those arising from the two-band model). Fig.~\ref{fig:hall}(b) shows the fits to $\rho_{xy}$ at \SI{150}{K} and \SI{2}{K}. Above $T_{CDW}$ the two-band model fits $\rho_{xy}$ well, but below $T_{CDW}$ the two-band fit loses quality, particularly in the low field regime. This disagreement is likely due to the contribution from an anomalous Hall effect~\cite{SVS_AHE, Mandrus_SVShall}. Fig.~\ref{fig:hall}(c) and Fig.~\ref{fig:hall}(d) present the carrier densities and mobilities, respectively, that are extracted from the two-band fitting. It should be noted that the quantitative values below $T_{CDW}$ should not be taken as exact due to the decrease in fit quality in $\rho_{xy}$. At all temperatures $n_h$ is greater than $n_e$ by several orders of magnitude. Notably, above $T_{CDW}$ $n_h$ decreases significantly, although $n_e$ is roughly constant, and below $T_{CDW}$ $n_h$ is roughly constant while $n_e$ grows by several times. Both $\mu_e$ and $\mu_h$ increase as temperature decreases which is typical for a metal.

\begin{figure*}
    \centering
    \includegraphics[width=0.9\textwidth]{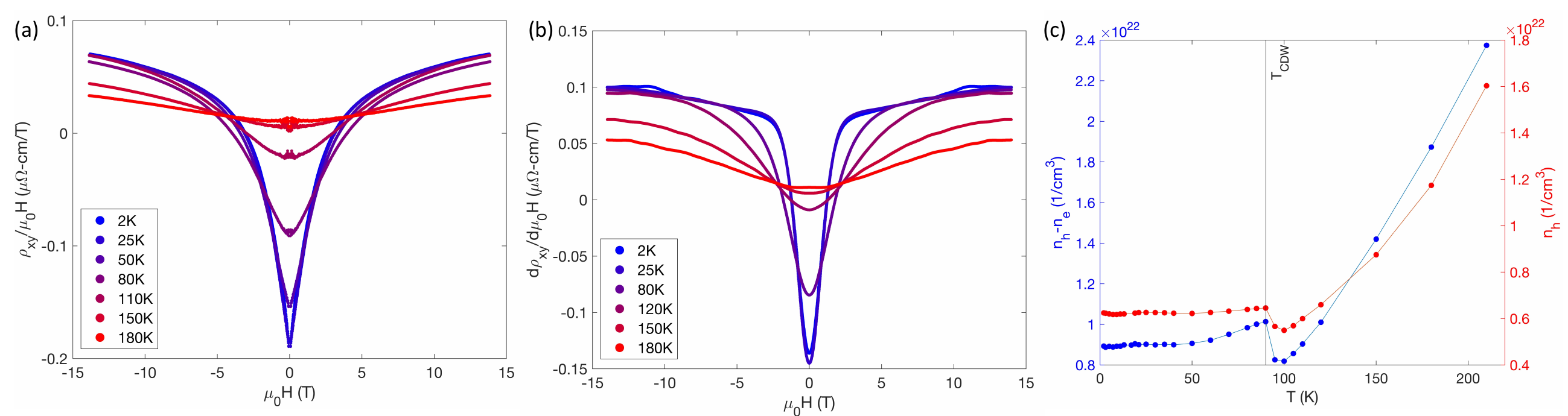}
    \caption{(a) $\rho_{xy}/\mu_0H$ as a function of $\mu_0H$ to evaluate the high field limit of the two-band model. Note that the curves are not fully saturated at \SI{14}{T}. (b) $\frac{d\rho_{xy}}{d\mu_0H}$ as a function of $\mu_0H$ to approximate the carrier concentration using a one-band model. (c) ${n_h-n_e}$ extracted from the high field limit of the two-band model and $n_h$ from the high field regime of a one-band model. While quantitatively different than the results presented in Fig.~\ref{fig:hall} for reasons described in the main text, all three of these analyses are qualitatively consistent.}
    \label{fig:alternativehall}
\end{figure*}

Due to the substantial deviation from the two-band model of the Hall data below $T_{CDW}$, we estimate the carrier density using the Hall data in the high field limit in two more ways to ensure our conclusions are robust. First, the Hall coefficient, $R_H = \frac{\rho_{xy}}{\mu_0H}$, should saturate to $\frac{1}{(n_h-n_e)e}$ at the high field limit where $e$ is the charge of an electron~\cite{Pippard_1988}. These data are shown in Fig.~\ref{fig:alternativehall}(a). The blue data in Fig.~\ref{fig:alternativehall}(c) present the extracted ${n_h-n_e}$ from the high field Hall coefficient. Second, the red data in Fig.~\ref{fig:alternativehall}(c) show $n_h$ using $\frac{d\rho_{xy}}{d\mu_0H}$ ($\mu_0H$ = \SI{14}{T}) $\approx \frac{1}{n_he}$. $\frac{d\rho_{xy}}{d\mu_0H}$ as a function of magnetic field is shown in Fig.~\ref{fig:alternativehall}(b). This method provides an estimate of carrier density in the presence of an anomalous Hall effect. While the three methodologies used to estimate the carrier densities vary quantitatively, they are all of the same order of magnitude and qualitatively consistent - above $T_{CDW}$ the number of holes decreases as a function of temperature, and below $T_{CDW}$ the number of holes is roughly constant.

\subsection{Magnetoresistance (MR) and Kohler's Rule Analysis}

\begin{figure*}
    \centering
    \includegraphics[width=0.9\textwidth]{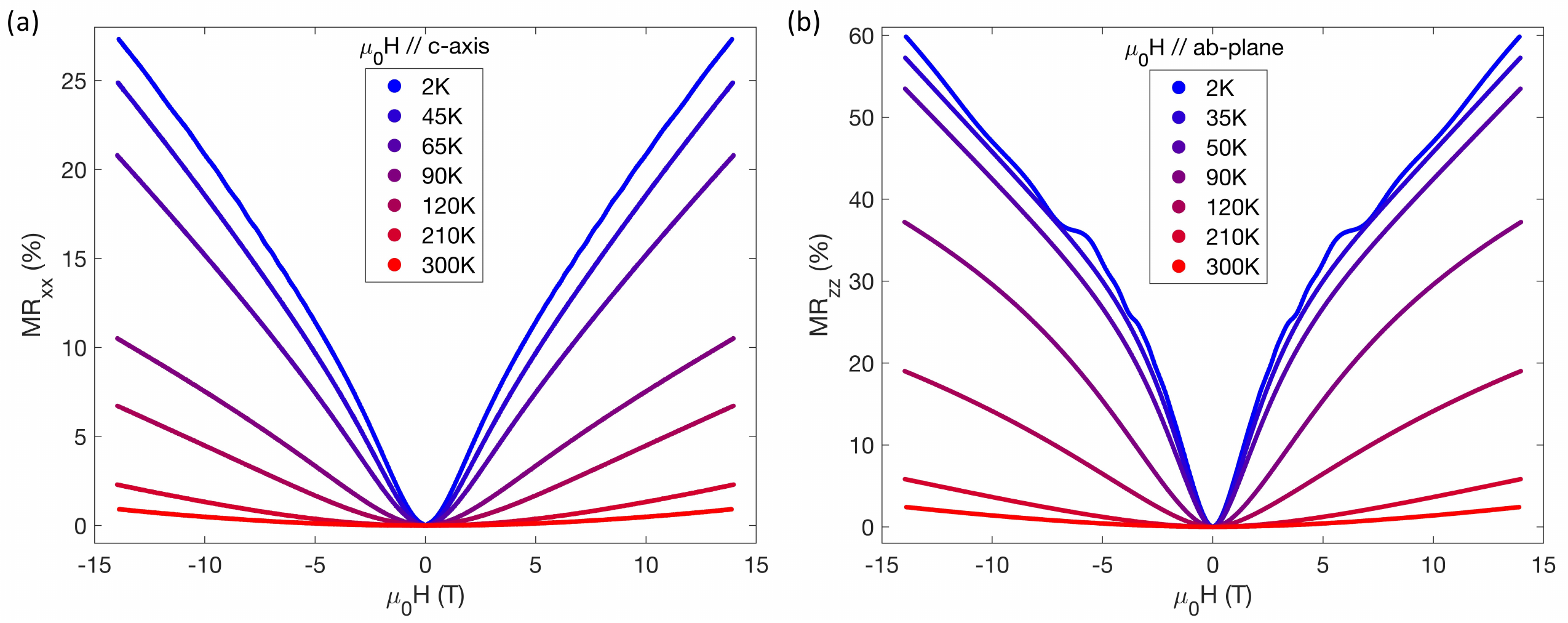}
    \caption{(a) $MR_{xx}$ as a function of magnetic field at various temperatures. (b) Magnetic field dependence of $MR_{zz}$ at several temperatures.}
    \label{fig:MR}
\end{figure*}

The ab-plane MR ($MR_{xx} = \frac{\Delta\rho_{xx}}{\rho_{xx}(\mu_0H=0)}*100\%$) with c-axis magnetic field at various temperatures is shown in Fig.~\ref{fig:MR}(a). This MR looks qualitatively similar to that observed in the \textit{A}\ch{V3Sb5} family in that the low field behavior shows a cusp at low temperatures and evolves to a more standard quadratic behavior at high temperatures~\cite{KVS_MR}.  At low temperatures quantum oscillations can be resolved once a background subtraction is performed.
The c-axis magnetoresistance ($MR_{zz}$) with a magnetic field in the ab-plane at several temperatures is presented in Fig.~\ref{fig:MR}(b). The $MR_{zz}$ of another sample was measured and was similar to the data presented here. While $MR_{zz}$ looks qualitatively similar to $MR_{xx}$, $MR_{zz}$ is about double the size at \SI{2}{K}. Also, at low temperatures quantum oscillations are observed much more prominently in $MR_{zz}$. These quantum oscillations, as well as accompanying electronic structure calculations, are discussed in the Supplemental Materials and are in agreement with a very recent study~\cite{SVS_AHE}. Overall, the quantum oscillations reveal three-dimensional Fermi pockets occupying less than a percent of the Brillouin zone with effective masses between 0.1 - 0.2 of the free electron mass. These pockets are consistent with the high mobility and low density of electron carriers extracted from the two-band Hall fitting. 

\begin{figure*}
    \centering
    \includegraphics[width=0.9\textwidth]{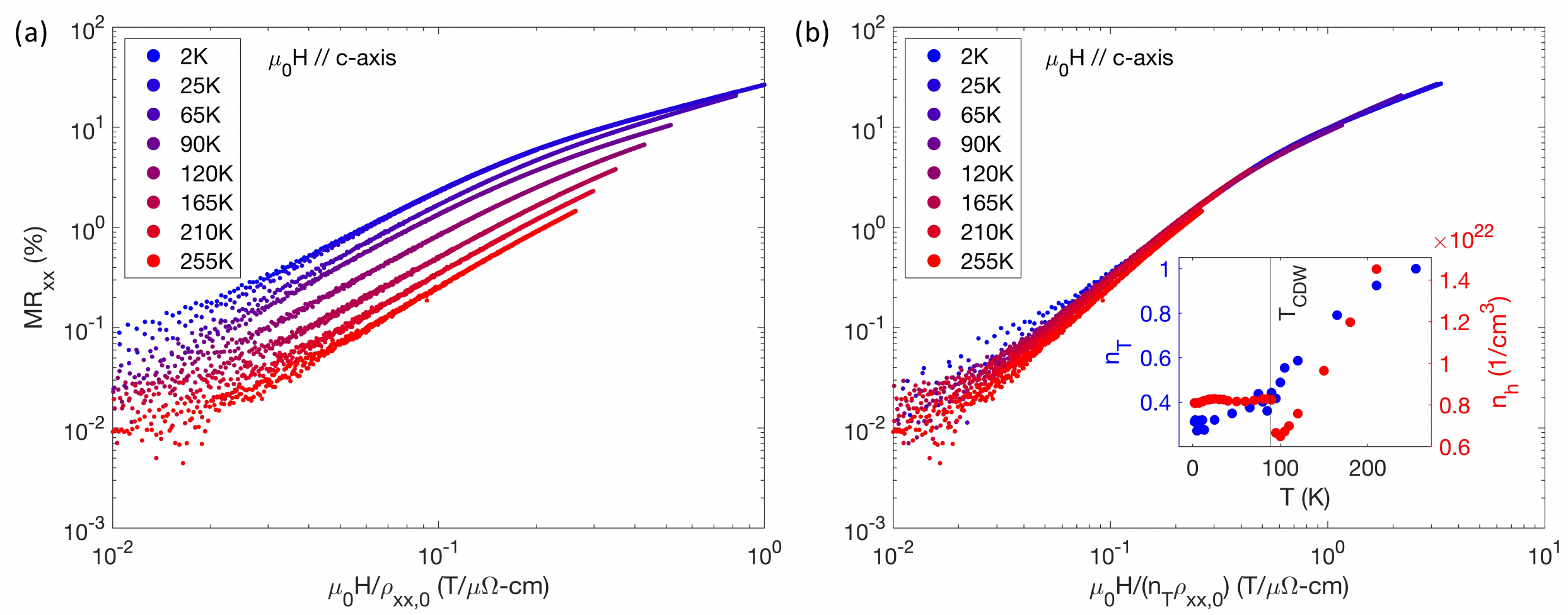}
    \caption{(a) $MR_{xx}$ as a function of $\mu_0H/\rho_{xx,0}$ using the data from Fig.~\ref{fig:MR}(a) on a log-log scale. Kohler's rule is violated as the data from different temperatures do not collapse onto each other. (b) Extended Kohler's rule applied to the same data as presented in (a) by plotting $MR_{xx}$ as a function of $\mu_0H/(n_T\rho_{xx,0})$ on a log-log scale. Inset: extracted $n_T$ as a function of temperature. $n_h$ extracted from the two-band Hall fitting as discussed in Section~\ref{subsection:Hall} is also plotted in the inset to highlight the similar temperature dependencies between $n_h$ and $n_T$.}
    \label{fig:EKR}
\end{figure*}

Kohler's rule of magnetoresistance~\cite{Kohlersrule} is violated in \ch{ScV6Sn6} as shown in Fig.~\ref{fig:EKR}(a) as  $MR_{xx}$ is not simply a function of $\mu_0H/\rho_{xx,0}$ where $\rho_{xx,0}$ is the zero-field resistivity. It should be noted that the data point spread at low fields and high temperatures is due to the small MR in this regime. The violation of Kohler's rule in \ch{ScV6Sn6} was also reported in a recent study~\cite{hiddenmagnetism_SVS}. Similar violations of Kohler's rule have been used as evidence of phase transitions~\cite{WTe2_phasetransition} or non-Fermi liquid behavior~\cite{cuprates_violationofkohler, fermiliquid_transport}. Recently, an extended Kohler's rule has been developed~\cite{extendedKohlersrule} in which the MR is expressed as a function of $\mu_0H/(n_T\rho_{xx,0})$ where $n_T$ describes the relative change in the carrier density. The extended Kohler's rule successfully explains the violation of conventional Kohler's rule by incorporating a temperature dependent carrier density, which could arise from a phase transition that partially gaps the Fermi surfaces or thermal excitations in topological semimetals where the Fermi energy is comparable to $k_BT$~\cite{extendedKohlersrule}. Here we apply this formula with $n_T$ fixed to be 1 at \SI{255}{K} to collapse the MR curves onto the linear part of the \SI{255}{K} MR (Fig.~\ref{fig:EKR}). Through this method, $n_T$ can be extracted as a function of temperature, and is presented in the inset of Fig.~\ref{fig:EKR}(b). Intriguingly, Kohler's rule is nearly followed below $T_{CDW}$ as evidenced by the nearly constant value of $n_T$, but above this temperature Kohler's rule is clearly violated. These results are consistent with those presented in Section~\ref{subsection:Hall} - $n_h$ extracted from two-band fitting is also plotted in the inset of Fig.~\ref{fig:EKR}(b) to show the similar behavior between the concentration of the dominant carriers (holes) and $n_T$, as both decrease above $T_{CDW}$ and become relatively constant below $T_{CDW}$. Analysis of $MR_{zz}$ data using the extended Kohler's rule yields similar results to that of $MR_{xx}$: above $T_{CDW}$ $n_T$ decreases quickly with decreasing temperature, and below $T_{CDW}$ $n_T$ changes far less drastically. These data are presented in the Supplemental Materials.

\section{Discussion}

Both Kohler's rule analysis and the Hall effect demonstrate a pronounced temperature dependence of the carrier density in \ch{ScV6Sn6}, decreasing by almost a factor of 2 from the value at 200K to the value just above the CDW transition temperature $T_{CDW} = 90K$. This decrease in carrier density is consistent with the insulating temperature dependence observed in interlayer resistivity. Interestingly, similar characteristics have also been observed in the Fe-based superconductors, such as \ch{BaFe2As2}. The temperature dependence of in-plane and interlayer resistivity, as well as the resistivity anisotropy ratio, exhibit remarkable resemblance between \ch{BaFe2As2} and \ch{ScV6Sn6}~\cite{BaCo122_caxis}. Additionally, in \ch{BaFe2As2} the Hall coefficient also shows a substantial increase with decreasing temperature above the spin density wave transition~\cite{CoBa122Hall1,CoBa122Hall2,PBa122Hall}. This is also seen in the pseudogap regime of the cuprate superconductors~\cite{badouxChangeCarrierDensity2016}. Another striking similarity can be observed in the magnetic susceptibilities of \ch{BaFe2As2} and \ch{ScV6Sn6} (presented in~\cite{SVSdiscovery}). In both materials, in addition to the drop below the phase transition due to gap formation, the susceptibility shows a linear increase with increasing temperature above the phase transition, which cannot be explained by either Pauli paramagnetic susceptibility or Curie Weiss susceptibility~\cite{FeSC_Sus,Zhang_2009}.         

In \ch{BaFe2As2}, the anomalous transport and magnetic properties observed above the transition temperature have been attributed to strong spin density wave fluctuations. However, in the case of \ch{ScV6Sn6}, the CDW transition is first-order, which could explain the nearly temperature independent carrier density below $T_{CDW}$, but contradicts with the existence of an extended fluctuation regime above $T_{CDW}$. Nevertheless, theoretical studies have suggested that, in addition to the long range $\sqrt{3}$ x $\sqrt{3}$ x 3 CDW that develops below $T_{CDW}$, there are several other nearly degenerate CDW instabilities associated with different ordering wave vectors~\cite{tanAbundantLatticeInstability2023a}. Furthermore, experimental evidence has shown the presence of a short-range $\sqrt{3}$ x $\sqrt{3}$ x 2 CDW well above $T_{CDW}$, which is suppressed by the $\sqrt{3}$ x $\sqrt{3}$ x 3 CDW through a first-order transition at $T_{CDW}$~\cite{competingCDW_SVS}. It is possible that the short-range CDW fluctuations are responsible for the anomalous decrease of carrier density and insulating interlayer resistivity observed in \ch{ScV6Sn6}.  


In conclusion, the transport behavior in the normal state of \ch{ScV6Sn6} is consistent with the formation of a pseudogap, which is likely arising from high temperature CDW fluctuations. We have also highlighted several similarities between \ch{ScV6Sn6} and Fe-based superconductors with pseudogaps above their ordering temperatures. Due to the high degree of tunability in the \textit{R}\textit{T}$_6$\textit{X}$_6$ family (\textit{R} = rare earth, \textit{T} = transition metal, \textit{X} = Si, Ge, Sn), \ch{ScV6Sn6} offers an exciting platform to study exotic electronic ordering in a kagome material.






\textit{Note:} 
During the preparation of this paper we became aware of a separate study which reported the two-band behavior of the Hall effect in \ch{ScV6Sn6}~\cite{Mandrus_SVShall}. They discovered high carrier density and low mobility holes and low carrier density and high mobility electrons, which broadly corroborates our findings.

\section{Methods}

Single crystals of \ch{ScV6Sn6} were grown using a flux method similar to one previously reported~\cite{SVSdiscovery}. Mixtures of Sc pieces (99.9\%), V pieces (99.9\%), and Sn shot (99.999\%) were loaded into Canfield crucible sets~\cite{Canfield2016} with atomic ratios 1:3:30, then vacuum-sealed in quartz tubes. These were heated to 1150$^\circ$C in 12 hours, held at this temperature for 15 hours, then cooled to 780$^\circ$C in 200 hours where the growths were decanted in a centrifuge to separate the excess flux from the single crystals. Dilute \ch{HCl} was used to etch the remaining flux from the surface of the crystals. The phase of the crystals was confirmed using energy-dispersive X-ray spectroscopy with a Sirion XL30 scanning electron microscope. The orientation of the crystallographic axes was determined using a Rigaku MiniFlex 600 system, with a Cu source and Hy-Pix 400MF 2D-detector.

Transport measurements were performed on samples that were polished and cut by a wire saw to be bar-shaped with dimensions roughly \SI{1}{mm} x \SI{0.4}{mm} x \SI{0.05}{mm} (in-plane current) or \SI{0.2}{mm} x \SI{0.15}{mm} x \SI{0.05}{mm} (out-of-plane current). Silver paste or two-part silver epoxy (H20-E) and gold wires were used to make 4 point and 5 point (Hall pattern) measurements. These measurements were performed in a Quantum Design Dynacool Physical Property Measurement System with standard lock-in techniques in temperatures ranging from \SI{1.7}{K} to \SI{300}{K} and in magnetic fields up to \SI{14}{T}. To eliminate any contributions from contact misalignment, the in-line and Hall resistivities were symmetrized and anti-symmetrized, respectively. For some of the measurements the samples were rotated \textit{in situ} using a Quantum Design in-plane rotator.

Density functional theory (DFT) calculations were performed using a full-potential linear augmented plane wave (FP-LAPW) method, as implemented in \textsc{wien2k}~\cite{WIEN2k}.
The primitive cell contains one formula unit, and experimental lattice parameters~\cite{SVSdiscovery} were adopted.
The generalized gradient approximation of Perdew, Burke, and Ernzerhof~\cite{perdew1996} was used for the correlation and exchange potentials.
To generate the self-consistent potential and charge, we employed $R_\text{MT}\cdot K_\text{max}=8.0$ with Muffin-Tin (MT) radii $R_\text{MT}=$ 2.4, 2.4, and 2.5 {a.u.}, for Sc, V and Sn, respectively.
The self-consistent calculations were performed with 490 $k$-points in the irreducible Brillouin zone (BZ). They were iterated until charge differences between consecutive iterations are smaller than $1\times10^{-3}e$ and the total energy difference lower than \SI{0.01}{mRy}.

 After obtaining self-consistent charge, band energies were calculated with a 64 $\times$ 64 $\times$ 33 fine k-mesh for the full Brillouin Zone (FBZ).  We employed FermiSurfer~\cite{fermisurfer} and SKEAF~\cite{skeaf} for visualizing FS and calculating de Haas - van Alphen (dHvA) frequencies, respectively.

\section*{Acknowledgements}

 This work is supported by the Air Force Office of Scientific Research under grant FA9550-21-1-0068, the David and Lucile Packard Foundation and the Gordon and Betty Moore Foundation’s EPiQS Initiative, grant no. GBMF6759 to J.-H.C.. This material is based upon work supported by the National Science Foundation Graduate Research Fellowship Program under Grant No. DGE-2140004. Any opinions, findings, and conclusions or recommendations expressed in this material are those of the authors and do not necessarily reflect the views of the National Science Foundation. This work was supported by the U.S.~Department of Energy, Office of Science, Office of Basic Energy Sciences, Materials Sciences and Engineering Division. Ames Laboratory is operated for the U.S.~Department of Energy by Iowa State University under Contract No.~DE-AC02-07CH11358.

\section*{Author Contributions}
J.M.D, E.R., O.P., and Z.L. conducted the transport measurements. Y.L. and L.K. performed the DFT calculations. Q.J. assisted with data analysis. J.M.D. and O.P. grew the samples. J.-H.C. oversaw the project. J.M.D., E.R., and J.-H.C. wrote the manuscript with input from all authors.

\section*{Competing Interests}
The authors declare no competing interests.

\clearpage

\bibliography{main}

\end{document}


\title{Supplementary Information for ``Pseudogap behavior in charge density wave kagome material ScV$_6$Sn$_6$ revealed by magnetotransport measurements"}

\author{Jonathan M. DeStefano}
\affiliation{Department of Physics, University of Washington, Seattle, WA 98112, USA}
\author{Elliott Rosenberg}
\affiliation{Department of Physics, University of Washington, Seattle, WA 98112, USA}
\author{Olivia Peek}
\affiliation{Department of Physics, University of Washington, Seattle, WA 98112, USA}
\author{Yongbin Lee}
\affiliation{Ames Laboratory, U.S. Department of Energy, Ames, Iowa 50011, USA}
\author{Zhaoyu Liu}
\affiliation{Department of Physics, University of Washington, Seattle, WA 98112, USA}
\author{Qianni Jiang}
\affiliation{Department of Physics, University of Washington, Seattle, WA 98112, USA}
\author{Liqin Ke}
\affiliation{Ames Laboratory, U.S. Department of Energy, Ames, Iowa 50011, USA}
\author{Jiun-Haw Chu}
\affiliation{Department of Physics, University of Washington, Seattle, WA 98112, USA}

\date{\today}

\maketitle

\section*{Supplementary Information}

\subsection{Quantum Oscillations and Electronic Structure}
\label{section:QOs}

\begin{figure*}
    \centering
    \includegraphics[width=0.9\textwidth]{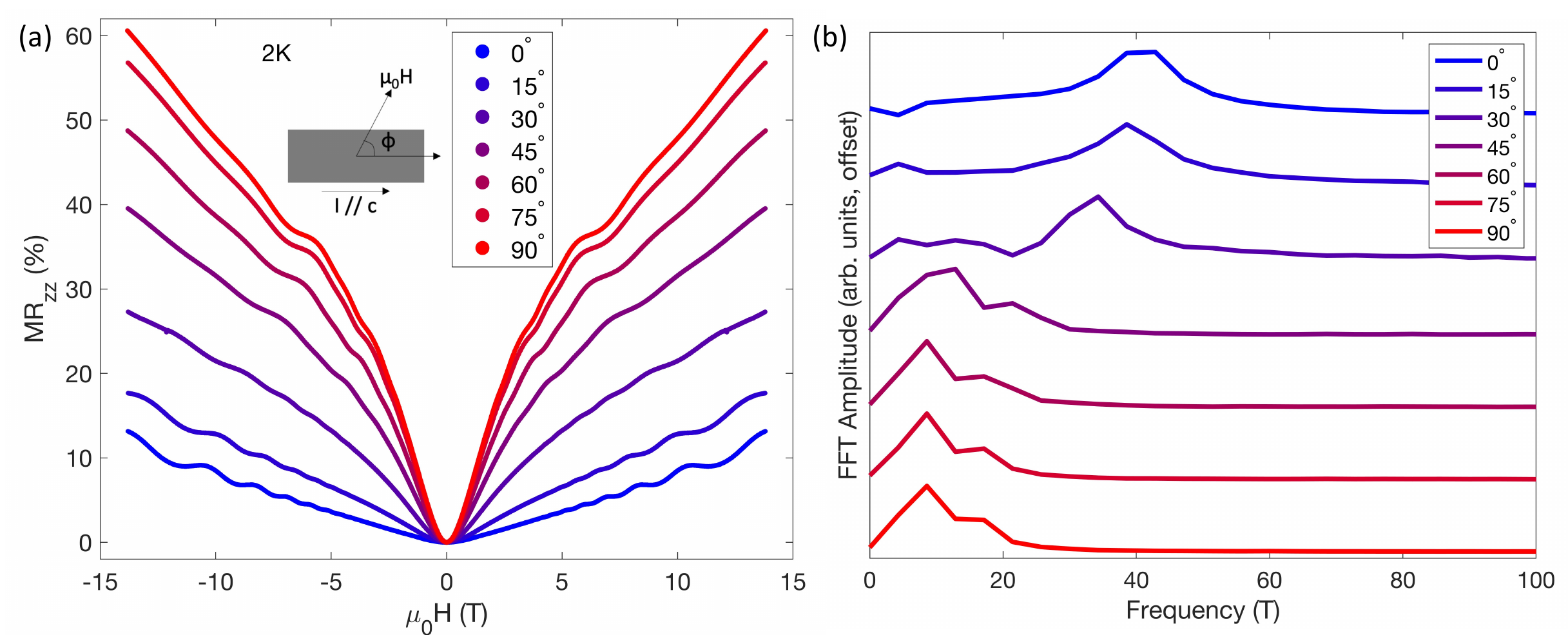}
\caption{(a) $MR_{zz}$ at \SI{2}{K} at several angles ranging from magnetic field along the c-axis to magnetic field in the ab-plane. (b) FFTs applied to the data shown in (a) from 3-\SI{10}{T}.}
    \label{fig:QOs_angles}
\end{figure*}

As observed in Fig. 4 of the main text, much more prominent quantum oscillations are observed in $\rho_{zz}$ than $\rho_{xx}$. As such, $\rho_{zz}$ is using for analysis of these oscillations. Supplementary Fig.~\ref{fig:QOs_angles}(a) presents $MR_{zz}$ at \SI{2}{K} as a function of magnetic field at select angles $\phi$ between the current and magnetic field which range from 0$^{\circ}$ (field along the c-axis) to 90$^{\circ}$ (field in the ab-plane). At a fixed magnetic field $\rho_{zz}$ increases monotonically as $\phi$ increases from 0$^{\circ}$ to 90$^{\circ}$.

The evolution of quantum oscillation frequencies with field angle is presented in Supplementary Fig.~\ref{fig:QOs_angles}(b). Fast Fourier transforms (FFTs) with a field range of  3-\SI{10}{T} were applied to the oscillatory parts of these data (a polynomial background was removed), revealing a single primary frequency of roughly \SI{43}{T} when field is along the c-axis, and two frequencies of \SI{9}{T} and \SI{17}{T} when field is in the ab-plane. 

\begin{figure*}
   \centering
    \includegraphics[width=0.9\textwidth]{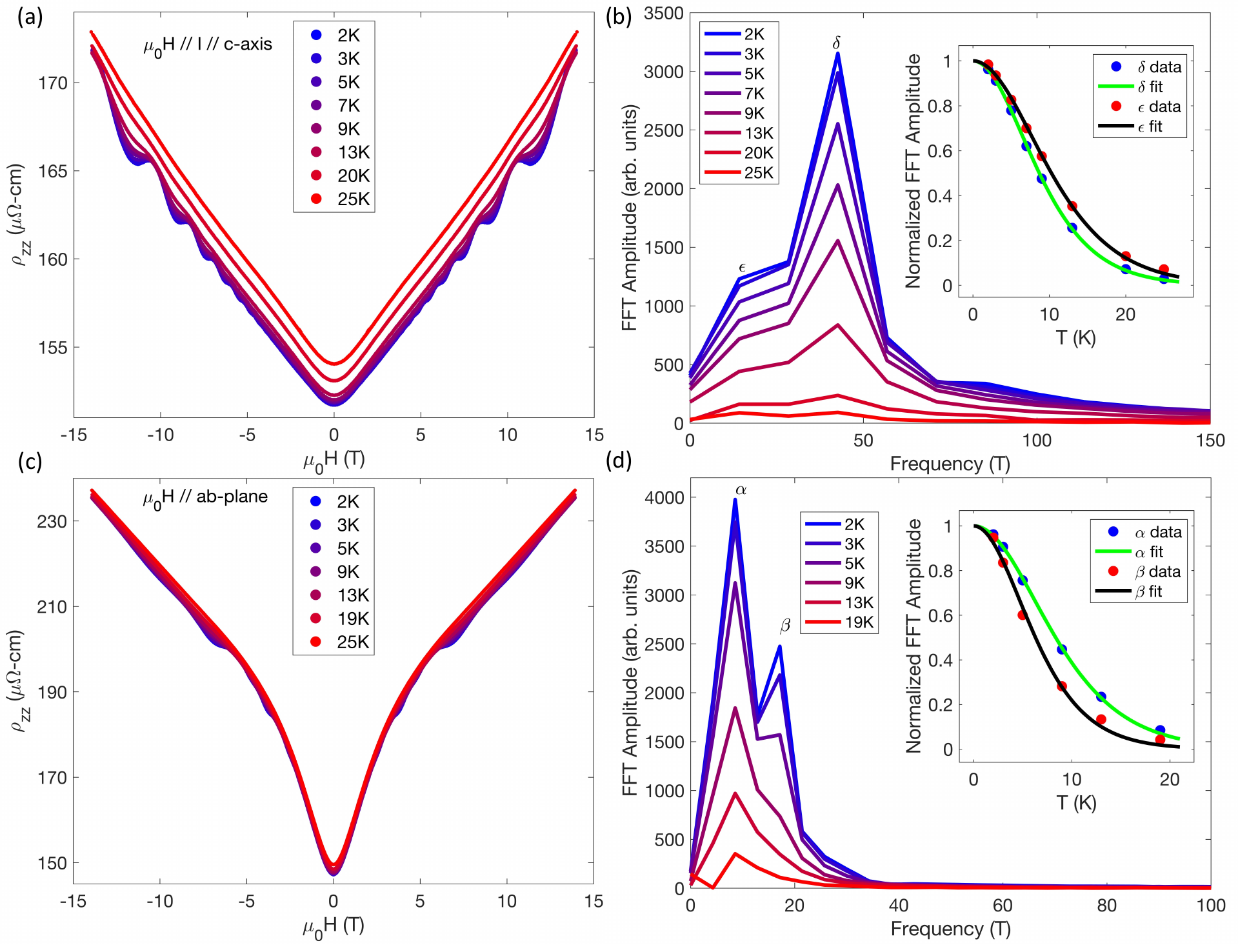}
\caption{$\rho_{zz}$ as a function of magnetic field at low temperatures with the field along the c-axis. (b) FFTs on the oscillatory parts of the data presented in (a) from 7-\SI{14}{T}. Inset: FFT amplitude of the major peak as a function of temperature fit to Eqn.~\ref{eqn:R_T}, revealing an effective mass of 0.18$m_o$. (c) $\rho_{zz}$ as a function of magnetic field at low temperatures with magnetic field in the ab-plane. (d) FFTs on the oscillatory parts of the data presented in (c) from 3-\SI{10}{T}. Inset: FFT amplitude of the $\alpha$ and $\beta$ peaks as a function of temperature fit to Eqn.~\ref{eqn:R_T}, revealing effective masses of 0.12$m_0$ and 0.15$m_0$, respectively.}
    \label{fig:QOs}
\end{figure*}

To extract the effective mass, the oscillations for field were measured as a function of temperature. Supplementary Fig.~\ref{fig:QOs}(a)and (c) display temperature dependence of $\rho_{zz}$ as a function of magnetic field along the c-axis and in-the ab-plane, respectively, with the FFT regions described in the caption of the figure. The FFT spectra are shown in Supplementary Fig.~\ref{fig:QOs}(b) and (d). The temperature dependence of the FFT peak amplitude is fitted with the damping term ($R_T$) in the Lifshitz-Kosevich formula~\cite{shoenberg_QOs}

\begin{equation}
    R_T = \frac{14.69m^*T/\mu_0H_{avg}}{\sinh{(14.69m^*T/\mu_0H_{avg})}}
    \label{eqn:R_T}
\end{equation}

where $\mu_0H_{avg}$ is the average of the magnetic field used in the FFT and $m^*$ is the effective mass. The inset of Supplementary Fig.~\ref{fig:QOs}(b) and (d) shows fitting which reveals an effective mass of 0.15$m_0$ and 0.18$m_0$ ($m_0$ is the free electron mass) for the \SI{14}{T} and \SI{43}{T} peaks when field is along the c-axis, and 0.12$m_0$ and 0.15$m_0$ for the \SI{9} and \SI{17}{T} peaks for field in-the ab-plane. Similar low frequency oscillations with small effective masses measured in other kagome systems including \textit{A}\ch{V3Sb5}~\cite{quantumtransport_CVS, KVS_MR, RVS_props} and $R$\ch{Mn6Sn6}~\cite{RareEarthEngineering_RMS}, which is likely originated from the same set of Dirac bands.

\begin{figure}
	\centering
	\begin{tabular}{c}
		\includegraphics[width=.5\linewidth,clip]{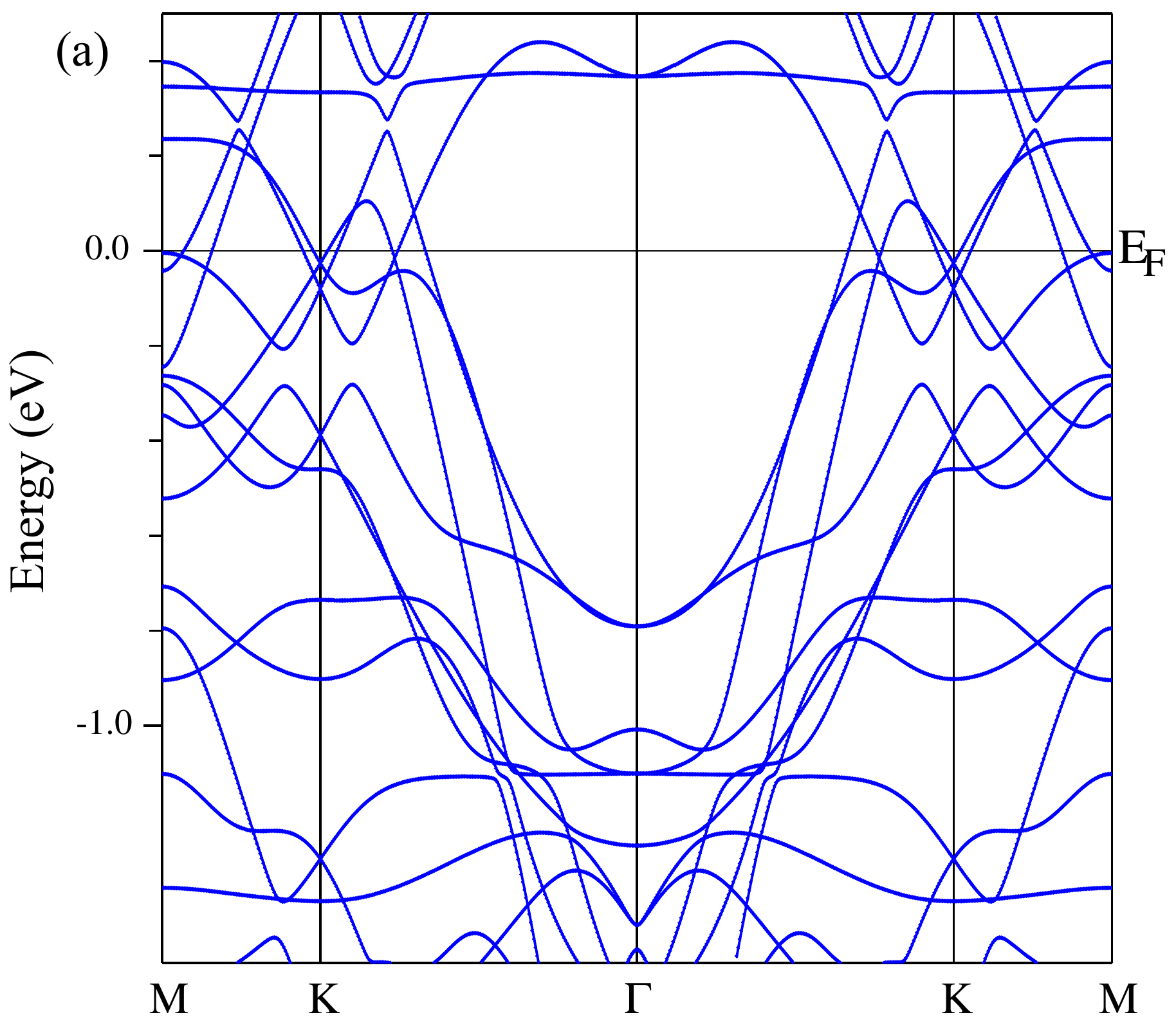}
        \includegraphics[width=.5\linewidth,clip]{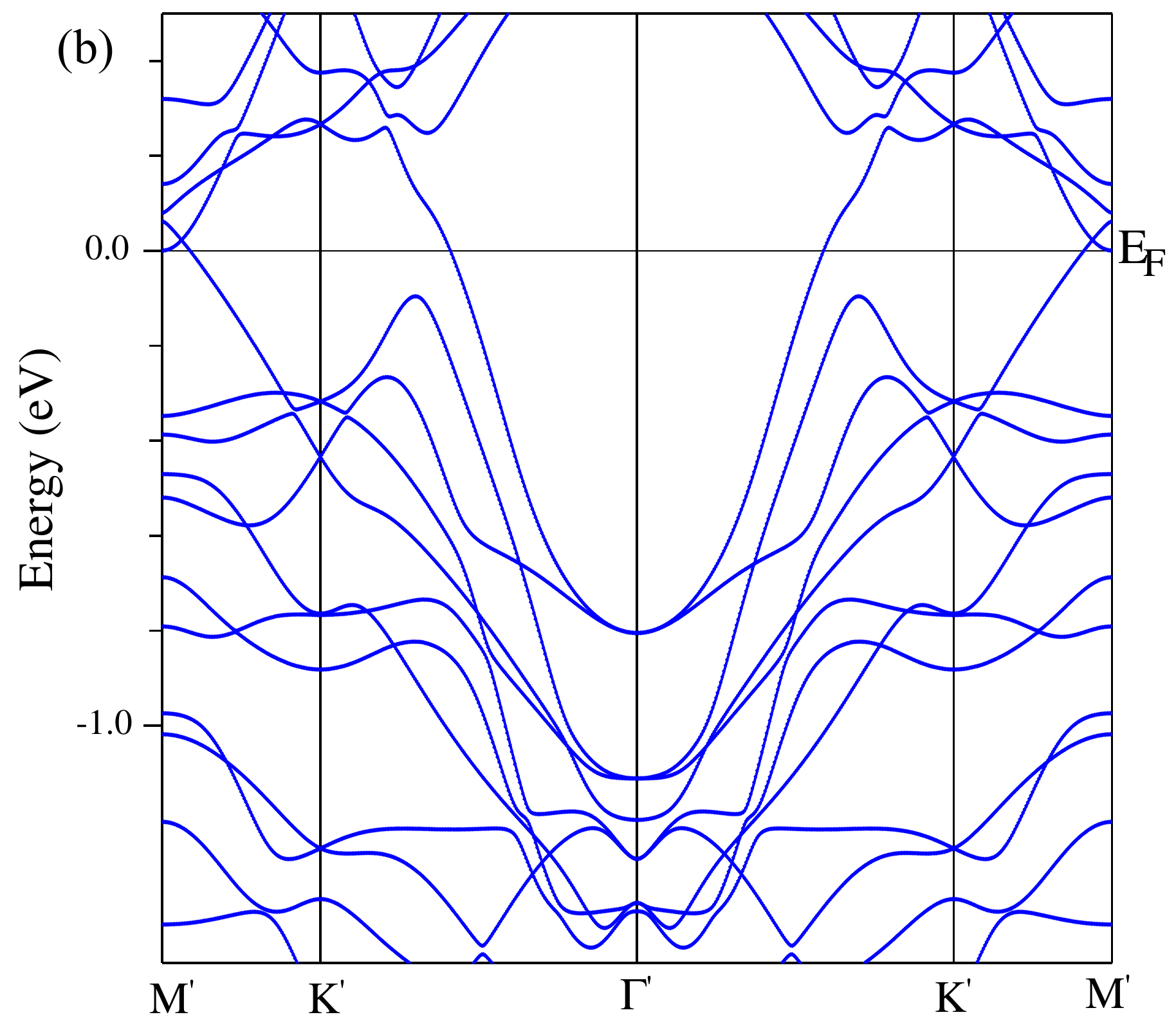}
	\end{tabular}%
	\caption{Non-spin-polarized band structures of ScV$_6$Sn$_6$ were calculated in two planes with $k_{z}$ = 0.0 (a) and 0.25 (b). In both planes, two bands intersect the E$_F$. 
		}
	\label{fig:band}
\end{figure}

\begin{figure}
	\centering
	\begin{tabular}{c}
		\includegraphics[width=.5\linewidth,clip]{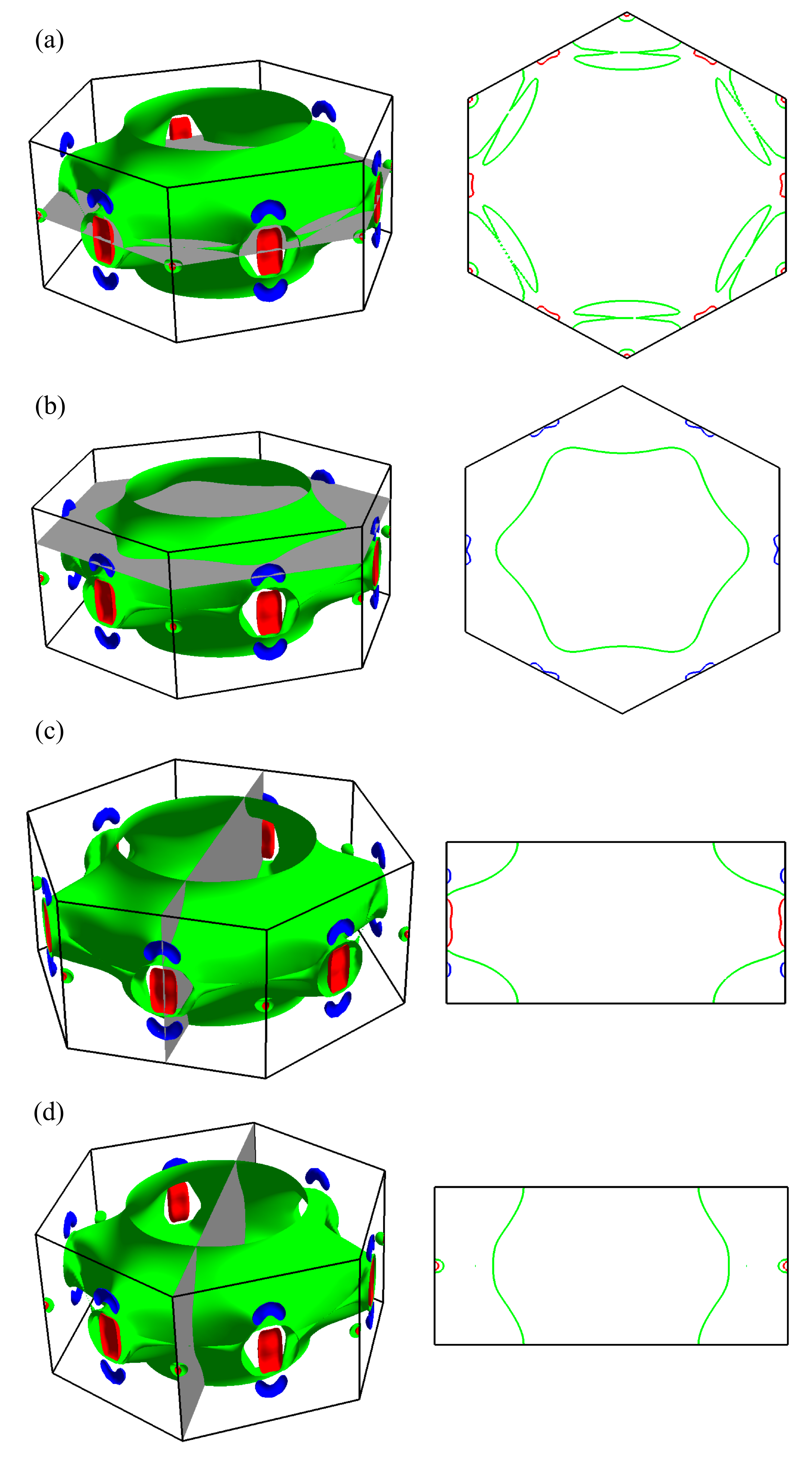} 
	\end{tabular}%
	\caption{3D Fermi surface (FS) plot along with cross sections for four chosen planes. (a) corresponds to the (0 0 1) plane with $k_z = 0.0$, (b) corresponds to the (0 0 1) plane with $k_z = 0.25$, (c) corresponds to the (1 1 0) plane passing through the $\Gamma$-point, and (d) corresponds to the (1 0 0) plane passing through the $\Gamma$-point. Different colors in the plot represent different bands.  Blue, green, and red correspond to band 1, 2, and 3, respectively.
	 }
	\label{fig:fs}
\end{figure}

To gain more insight, the band structure and Fermi surfaces (FS) of ScV$_6$Sn$_6$ were calculated using the density functional theory based on the non-distorted crystal structure. Supplementary Fig.~\ref{fig:band} displays non-spin-polarized band structures in two planes with $k_{z}$ = 0.0 (a) and 0.25 (b). In both planes, two bands intersect the E$_F$. On the $k_{z}$ = 0.0 plane, there are one electron pocket around the $M$-point and two electron pockets around the $K$-point. However, on the $k_{z}$ = 0.25 plane, there is only one hole pocket around the $M'$-point, and there is no pocket around the $K'$-point.
Fig.~\ref{fig:fs} presents a 3D Fermi Surface (FS) plot along with cross sections for four chosen planes. (a) corresponds to the (0 0 1) plane with $k_z = 0.0$, (b) corresponds to the (0 0 1) plane with $k_z = 0.25$, (c) corresponds to the (1 1 0) plane passing through the $\Gamma$-point, and (d) corresponds to the (1 0 0) plane passing through the $\Gamma$-point. Different colors in the plot represent different bands. Blue, green, and red correspond to band 1, 2, and 3, respectively.

In the FS plot, three bands contribute to the construction of the Fermi surface. The calculation reveals a large open Fermi surface and several small pockets at the boundary of the first Brillouin zone.

\begin{figure}
	\centering
	\begin{tabular}{c}
		\includegraphics[width=.5\linewidth,clip]{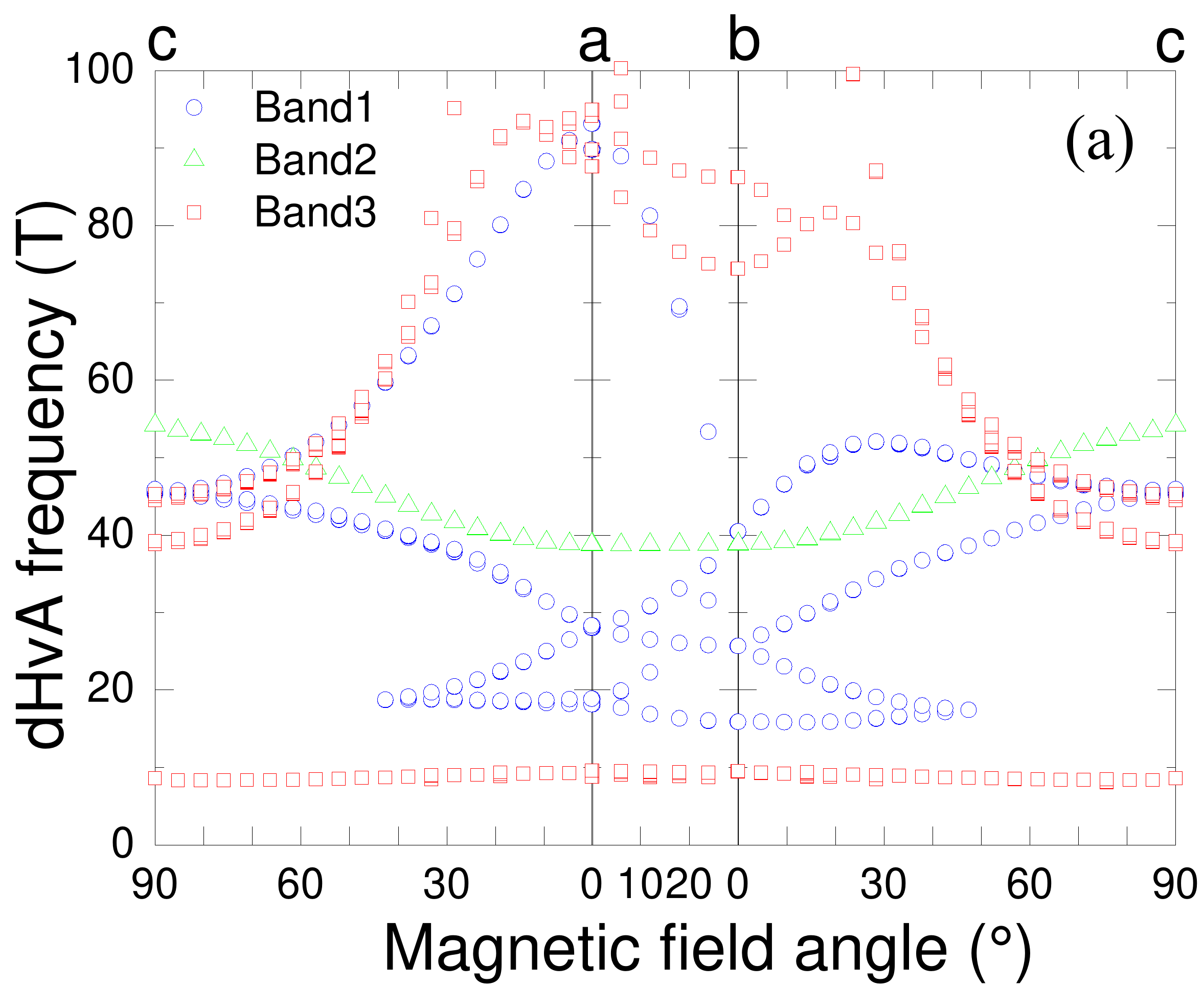} \\
		 \includegraphics[width=.5\linewidth,clip]{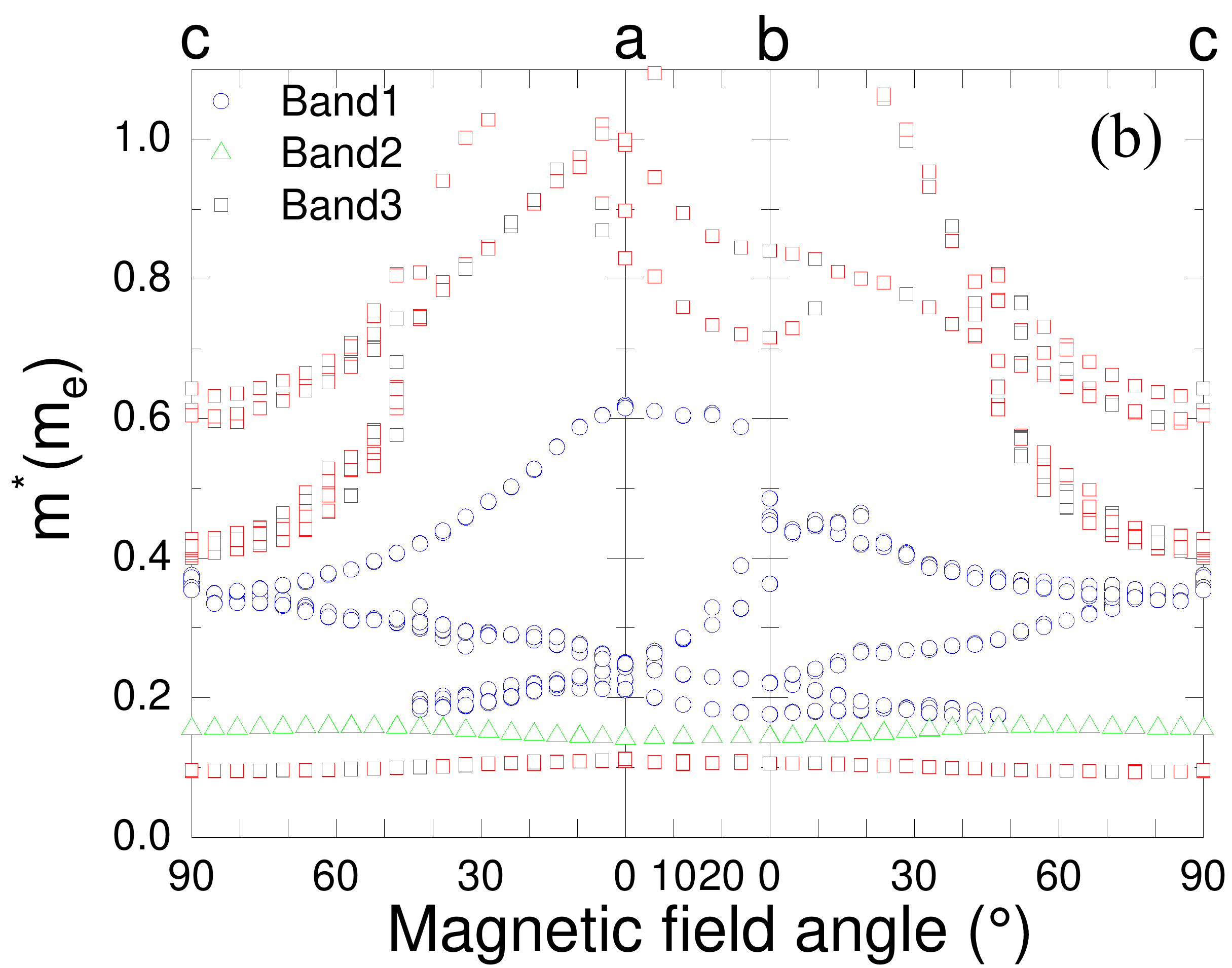}
	\end{tabular}%
	\caption{We have calculated the dHvA frequencies (a) and the effective mass (b) of the corresponding orbitals. Since our focus is on smaller pockets at the zone-boundary, we are presenting data that are smaller than 100T. For the c-axis, the external magnetic field is applied along the hexagonal axis. For the a-axis, it is parallel to the $\Gamma$-$M$ direction, and for the b-axis, it is along the $\Gamma$-$K$ direction. Color scheme is consistent with Fig.~\ref{fig:fs}.
		}
	\label{fig:freq}
\end{figure}

We also calculated the de Haas-van Alphen (dHvA) frequencies (Fig.~\ref{fig:freq}(a)) and the effective mass (Supplementary Fig.~\ref{fig:freq}(b)) of these small orbitals. For the c-axis, the external magnetic field is applied along the hexagonal axis. For the a-axis, it is parallel to the $\Gamma$-$M$ direction, and for the b-axis, it is along the $\Gamma$-$K$ direction.
Supplementary Table~\ref{tbl:eff-mass} displays dHvA frequencies (for frequency $<$ \SI{100}{T}) and effective masses with the magnetic field applied along $c$, $a$ and $b$-directions. 

Despite using the non-distorted crystal structure in our calculations, the calculated angle dependence of the frequency and effective mass of these smaller pockets are in broad agreement with the experimental observations.  For example, the observed \SI{43}{T} peak with an effective mass of 0.18$m_0$ for field along the c-axis may correspond to the electron pockets from band 2 (green) located at K point, which has a frequency of \SI{54}{T} and an effective mass of 0.16$m_0$ for field along the c-axis.  The observed \SI{9}{T} peak with an effective mass of 0.12$m_0$ for field in the ab-plane agrees with the highly isotropic electron pockets from band 3 (red) centered at K points, which also has a frequency $\approx$ \SI{9}{T} and an effective mass of 0.12$m_0$ at all angles. The existence of these tiny electron pockets with small effective mass is consistent with the low density and high mobility electron carriers extracted from the two band Hall fitting. Assuming an ellipsoid-shaped pocket with frequencies of \SI{43}{T} with field along the c-axis and \SI{17}{T} with field in-plane yields an estimated carrier concentration of $\approx6.3$x$10^{17}$ cm$^{-3}$.  Considering that there are multiple identical pockets in the 1st BZ, this estimation gives a value of roughly the same order of magnitude as the extraction electron carrier density from two-band Hall fitting presented in Fig. 2(c) of the main text.



\begin{table}
	\caption{Calculated dHvA frequencies (frequency $<$ \SI{100}{T}) and effective masses with external magnetic field (B-field) applied in three different directions. For the $c$-direction, B-field is applied along the hexagonal axis. For the $a$-direction, it is along the $\Gamma$-$M$ direction and for the $b$-direction, it is parallel to the $\Gamma$-$K$ direction.  The effective masses are given in units of electron mass. 
	}
	\label{tbl:eff-mass}
	\bgroup
	\def\arraystretch{1.3}
	\begin{tabular*}{\linewidth}{cc@{\extracolsep{\fill}}ccccccrrrrrrlrcccr}
		\hline\hline
		\multirow{2}{*}{B-field} &  & \multicolumn{3}{c}{dHvA frequency (T)} & & \multicolumn{3}{c}{effective mass (m$_e$)}  \\  \cline{3-5}  \cline{7-9} 
		&  & Band1 & Band2 & Band3  & & Band1 & Band2 & Band3  \\
		\hline
		c     &  & 45.6  &54.2  & 8.6     &  & 0.37 & 0.16 & 0.10  \\
		&  &       &      & 38.8    &  &      &      & 0.61  \\
		&  &       &      & 45.0    &  &      &      & 0.41  \\      
		\hline
		a     &  & 18.8  &38.8  & 8.8     &  & 0.23 & 0.14 & 0.11  \\
		&  & 28.1  &      & 87.6    &  & 0.25 &      & 0.99  \\
		&  & 89.8  &      & 94.9    &  & 0.62 &      & 0.83  \\
		\hline
		b     &  & 15.9  &38.8  & 9.5     &  & 0.18 & 0.14 & 0.11  \\
		&  & 25.7  &      & 74.4    &  & 0.22 &      & 0.84  \\
		&  & 40.5  &      & 86.2    &  & 0.45 &      & 0.72  \\          
		
		\hline\hline
	\end{tabular*}
	\label{tbl:ms_morb_mtot}
	\egroup
\end{table}

\subsection{c-axis Kohler's Rule Analysis}
\label{section:caxisKohler}

\begin{figure*}
    \centering
    \includegraphics[width=0.9\textwidth]{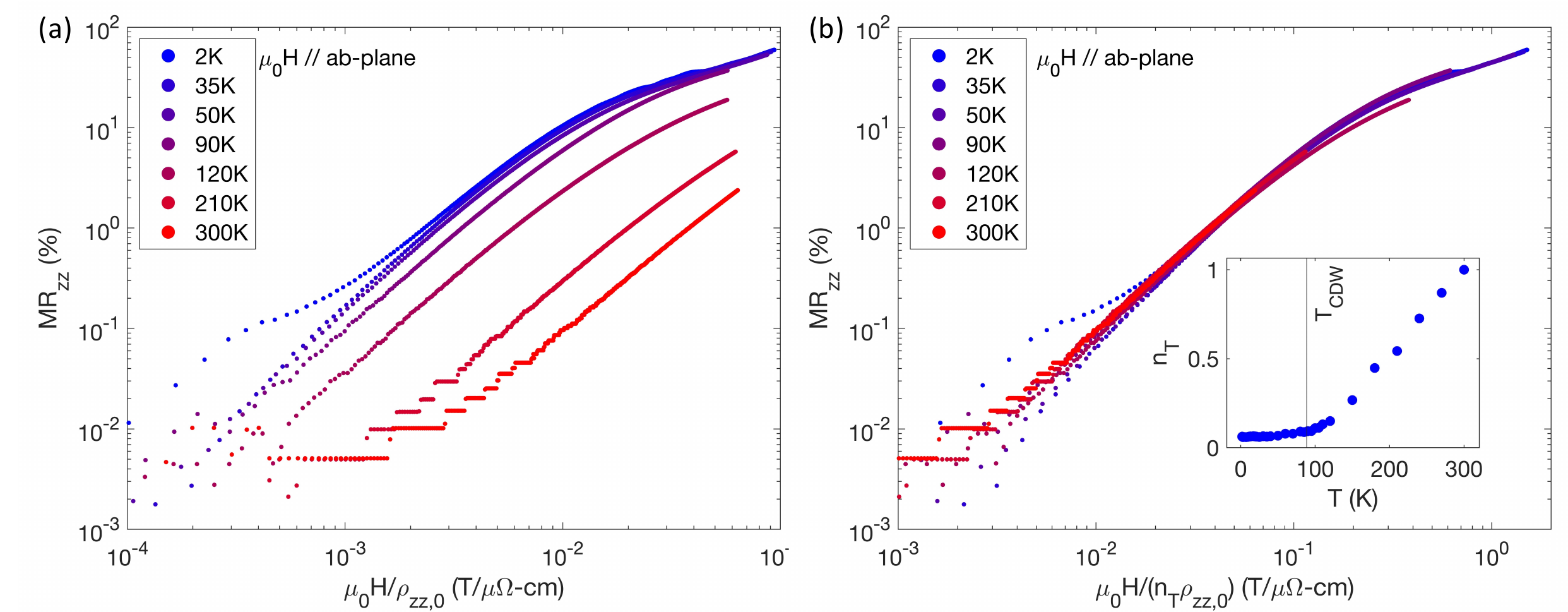}
    \caption{(a) $MR_{zz}$ as a function of $\mu_0H/\rho_{zz,0}$ using the data from Fig. 4(b) from the main paper on a log-log scale. Kohler's rule is violated as the data from different temperatures do not collapse onto each other. (b) Extended Kohler's rule applied to the same data as presented in (a) by plotting $MR_{zz}$ as a function of $\mu_0H/(n_T\rho_{zz,0})$ on a log-log scale. Inset: extracted $n_T$ as a function of temperature.}
    \label{fig:EKRzz}
\end{figure*}

Kohler's rule of magnetoresistance is violated in c-axis transport of \ch{ScV6Sn6} as shown in Fig.~\ref{fig:EKRzz}(a) as  $MR_{zz}$ is not simply a function of $\mu_0H/\rho_{zz,0}$ where $\rho_{zz,0}$ is the zero-field resistivity. Fig.~\ref{fig:EKRzz}(b) shows the same data with extended Kohler's rule applied as described in the main text. The inset of this figure shows the extracted $n_T$. Similar to in-plane transport, this inter-plane transport shows a drastic decrease of $n_T$ with decreasing temperature above $T_{CDW}$ and below $T_{CDW}$ $n_T$ is nearly constant.




\clearpage

\bibliography{main}